\documentstyle[12pt]{article}
\newcommand{\dv}{\hbox {$\Delta \rm V^{TO}_{HB}$}}
\newcommand{\nli}{\log {\rm N} ({\rm Li})}
\newcommand{\msun}{\hbox {${\rm M}_{\odot}$}}
\newcommand{\li}{\hbox {$^7$Li}}
\newcommand{\teff}{\hbox {$\rm T_{eff}$}}
\newcommand{\ea}{\hbox {\it et al.}}
\newcommand{\mvrr}{\hbox {$\rm M_v(RR)$}}
\newcommand{\feh}{\hbox{$ [{\rm Fe}/{\rm H}]$}}
\newcommand{\la}{\mathrel{\hbox{\rlap{\hbox{\lower4pt\hbox{$\sim$}}}\hbox{$<$}}}}
\newcommand{\ga}{\mathrel{\hbox{\rlap{\hbox{\lower4pt\hbox{$\sim$}}}\hbox{$>$}}}}
\newcommand{\bull}{\vrule height .9ex width .8ex depth -.1ex}

\title{\bf Halo Star Evolution}
\author{Brian Chaboyer
\vspace{1cm}\\
\normalsize Canadian Institute for Theoretical Astrophysics,
Toronto, Ontario, Canada}
\date{\mbox{}}
\begin{document}
\maketitle
%
% AND DEFINE OUR MACROS FOR THE REFERENCE LIST
% I.E \beginrefer \refer and \endrefer
%
\def\beginrefer{\section*{References}%
\begin{quotation}\mbox{}\par}
\def\refer#1\par{{\setlength{\parindent}{-\leftmargin}\indent#1\par}}
\def\endrefer{\end{quotation}}
%
% BEGIN THE ABSTRACT CHAPTER WITH \noindent\small, ENCLOSE IT IN A GROUP
% AND BOLDFACE THE TITLE.
%
{\noindent\small{\bf Abstract:}

The low mass, low metallicity stars which make up the halo are
probably the simplest of all stars to model. However, there are
several puzzling discrepancies between theory and observations.  In
this review, I will discuss a few problems which point to the need for
improved stellar evolution models of halo stars.  Observations of the
surface abundance of $^7$Li provide a sensitive diagnostic of the
internal structure of cool stars.  Current stellar evolution models do
not match the observed $^7$Li abundance patterns, suggesting that the
input physics and/or the assumptions used in constructing the models
are in need of revision.  It appears that all halo stars have suffered
some $^7$Li depletion, implying that the primordial $^7$Li abundance
is higher than that presently observed in hot halo stars.
Observations of abundances of various elements in globular cluster
giant branch stars have suggested for some time now that some form of
deep mixing, which is not present in theoretical models, occurs in
halo stars.  The driving mechanism for this mixing, and its
incorporation into stellar models remain one of the key problems in
stellar modeling.  Current theoretical isochrones are able to provide
a good match to observed colour-magnitude diagrams.  However, there is
some evidence that the theoretical luminosity functions are in
disagreement with observations.  This is an area which requires
further study, as it suggests that the relative main sequence/giant
branch lifetimes predicted by the models are incorrect.  A discussion
of some of the uncertainties involved in determining the ages of
globular clusters is presented.  The absolute ages of globular
clusters provide a lower bound to the age of the universe, and so are
of great interest to cosmologists.  Unfortunately, present
uncertainties in stellar models lead to a rather large range in the
inferred ages of globular clusters of 11 -- 18 Gyr.  }

%
% NOW COMES THE MAIN BODY OF THE ARTICLE
%
\section{Introduction}
Halo stars are metal-poor ($\feh \la -1.0$) stars which typically have
high velocity, randomly oriented orbits.  As such, they were among the
first objects formed in the Milky Way and so provide important
constraints on the early formation history of the Milky Way, and
galaxy formation in general.  In addition, the fact that
halo stars where formed soon after the big bang implies that
determining a reliable estimate for their age yields a reasonable
minimum estimate for the age of the universe.  Finally, the primordial
abundance of $^7$Li may be estimated by observations of halo stars
coupled with stellar evolution calculations.  As $^7$Li is produced
via big bang nucleosynthesis, an accurate determination of its
primordial abundance provides a diagnostic of the conditions in the
very early universe.

\vspace*{-24cm}
{\tiny\bull\quad ``Stellar Evolution: What Should Be Done'';
32$^{\rm st}$ Li\`ege\ Int.\ Astroph.\ Coll., 1995\quad\bull\hfill}
{CITA--95--14}\\
{\tiny\hspace*{17.5pt}\bull\quad eds.~A.~Noels, M.~Gabriel, N.~Grevesse \&
P.~Demarque (Li\`ege: Institut d'Astrophysique) \quad\bull }\hfill

\clearpage

The importance of halo stars to astrophysics has long been recognized,
and theoretical stellar evolution models of halo stars have been
calculated for many years.  A major uncertainty in these early stellar
models, which remains today, is the treatments of convection.  A
detailed understanding of the structure in the transition region
between radiative and convective regions continues to be one of the
key quests in stellar astrophysics.  Fortunately, for the main
sequence and red giant branch stars which are in the halo, convection
is only important in the outer layers.  In this review, I will
concentrate on these early stages of stellar evolution, emphasizing
the problems which are encountered with current stellar evolution
models of halo stars when they are compared to observations.  The more
advanced stages of stellar evolution are covered by Lattanzio and
Dorman in this volume.  Due to space limitations, I have chosen to
focus on four areas of halo star stellar evolution research.  Section
2 discusses the importance of $^7$Li for constraining stellar
evolution models, and big bang nucleosynthesis.  The long standing
difficulty in understanding the abundance anomalies observed in red
giant branch stars is highlighted in Sect.~3.  Section 4 will assess
the current status of theoretical isochrone and luminosity function
calculations.  A discussion of the uncertainties in globular cluster
ages estimates is presented in Sect.~5.  A brief summary is presented
in Sect.~6.

\section{$^7$Li}
Observations of lithium\footnote{In stars, Li may exist
in two different isotopes: $^6$Li and \li.  It is extremely difficult
observationally to distinguish between $^6$Li and $^7$Li, thus the
observations typically determine the total Li present in a star.
However, $^6$Li is even more fragile than \li ~and so is likely to be
destroyed in most stars.  In addition, Smith, Lambert \& Nissen (1992)
determined the abundance ratio of $^6$Li/\li ~to be $0.05\pm 0.02$ in
the halo star HD 84937.  It appears that the total Li content in a
star is dominated by \li.  Thus, we will assume that the Li
observations actually measure \li.}  provide a critical test of
stellar models, for \li ~is destroyed at temperatures around $2.6\times
10^6\,$K, which is located in the outer envelope of most halo stars.  Hence,
its observed surface abundance can be dramatically affected by small
changes in the convection zone depth, or small amounts of mixing in
the stellar radiative regions below the surface convection zone.  In
addition to its importance for testing stellar models, \li ~is
interesting because it was produced in significant quantities during
the big bang.  Thus, an accurate determination of the primordial
abundance of \li ~serves as a test of big bang nucleosynthesis
(eg.~Smith, Kawano \& Malaney 1993).  For
these reasons, there have been a large number of theoretical and
observational studies of \li ~in halo stars.

Observational studies have revealed a remarkably uniform \li
{}~abundance of $\nli \simeq 2.2$ (where $\nli \equiv \log ({\rm Li/H})
+ 12$) over a wide range of effective temperature and metallicity
($\feh \la -1.5$, $\teff \ga 5700\,$K; Spite \& Spite 1982;
Thorburn 1994).  This \li ~plateau provides a severe constraint for
any mixing mechanism which operates in the radiative regions of a
star, for such a mixing mechanism must provide a nearly uniform
depletion of \li ~over a wide range of effective temperatures and
metallicities.  This constraint is often assumed to be so strong as to
preclude any \li ~depletion in halo stars, and so the observed
abundance is assumed to represent the primordial abundance (eg.~Walker
\ea ~1991).  However, as shown by Boesgaard (1991) {\em \li ~has been
depleted in all Pop I stars which possess a surface convection zone}.
As the structure of the plateau stars is similar to the Pop I stars,
this is a strong argument in favour of \li ~depletion occurring in the
plateau stars.  In addition, we note that the study of Thorburn
(1994) has revealed that about 5\% of the halo stars with effective
temperatures in the plateau regions have \li ~abundances which are
depleted by at least an order of magnitude.  This clearly indicates
that large \li ~depletions occur in some `plateau' stars, and presents
a further challenge to theoretical models.

\begin{figure}
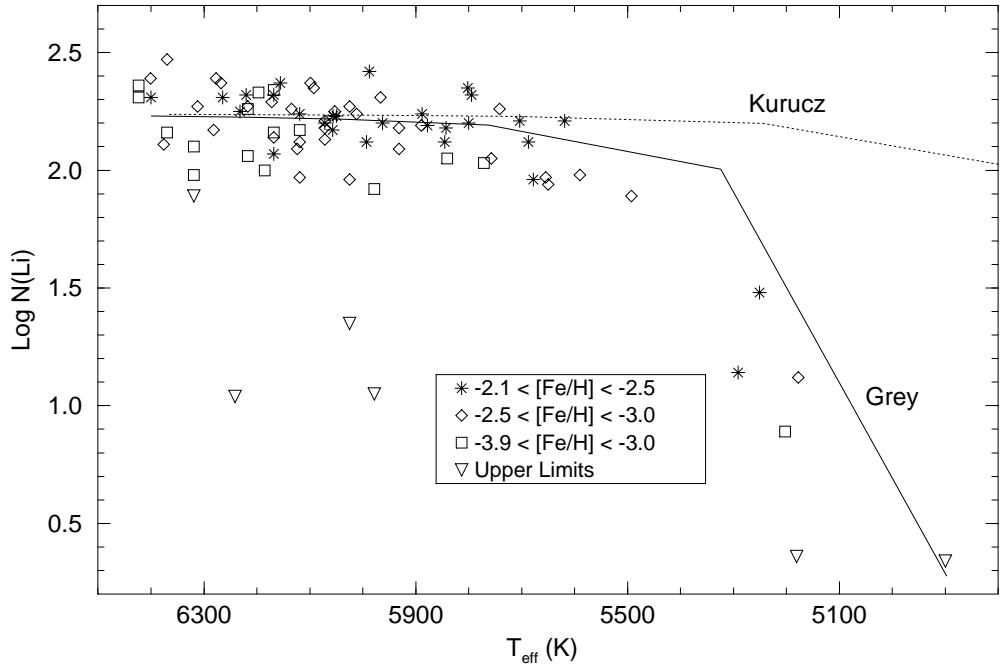

%\centerline{\psfig{figure=haloatm.ps,height=9.0cm}}
\vspace*{9cm}
\caption
{Comparison of Kurucz and grey atmosphere $^7$Li isochrones to the
data from Thorburn (1994). The isochrones shown have an age of 18 Gyr
and an overshoot of $0.02$ pressure scale heights.  The initial \li
{}~abundance was taken to be 2.25.}
\label{kurli}
\end{figure}

For halo stars which are cooler than the plateau ($\teff \la 5600\,$K),
significant amounts of \li ~depletion has occurred.  This provides a
further test for stellar models, for the amount of \li ~depletion in
cool star halo models is very sensitive to the depth of the surface
convection zone.  Theoretical models predict that this \li ~depletion
occurs largely on the pre-main sequence, where the convection zone
has its greatest depth.  Figure \ref{kurli} compares \li ~depletions predicted
by standard stellar models with the latest available input physics
(opacities, nuclear generation rates, etc.) from Chaboyer \& Demarque
(1994) to the observations of Thorburn (1994).  Two different
theoretical models are shown; one set uses a grey model atmosphere for
the surface boundary condition while the other set uses the Kurucz (1992)
model atmospheres.  These models match the bulk of the plateau star
observations, though they are unable to account for the depleted plateau
stars.  However, the models which use the Kurucz (1992) atmospheres
predict very little \li ~depletion in cool stars, in clear contrast to
the observations.  Thus, stellar models with the best available input
physics are unable to match the observed \li ~abundances in halo
stars.  This is also true for models which incorporate extra mixing in
the radiative regions (Chaboyer \& Demarque 1994). Note that
models which employ the grey model atmospheres do a better job of
matching the observations, but still predict too little \li ~depletion
in the cool stars. It is clear that improvements in the basic physics
(model atmospheres, opacities, etc.) used in stellar models of halo
stars are required.

\begin{figure}[t]
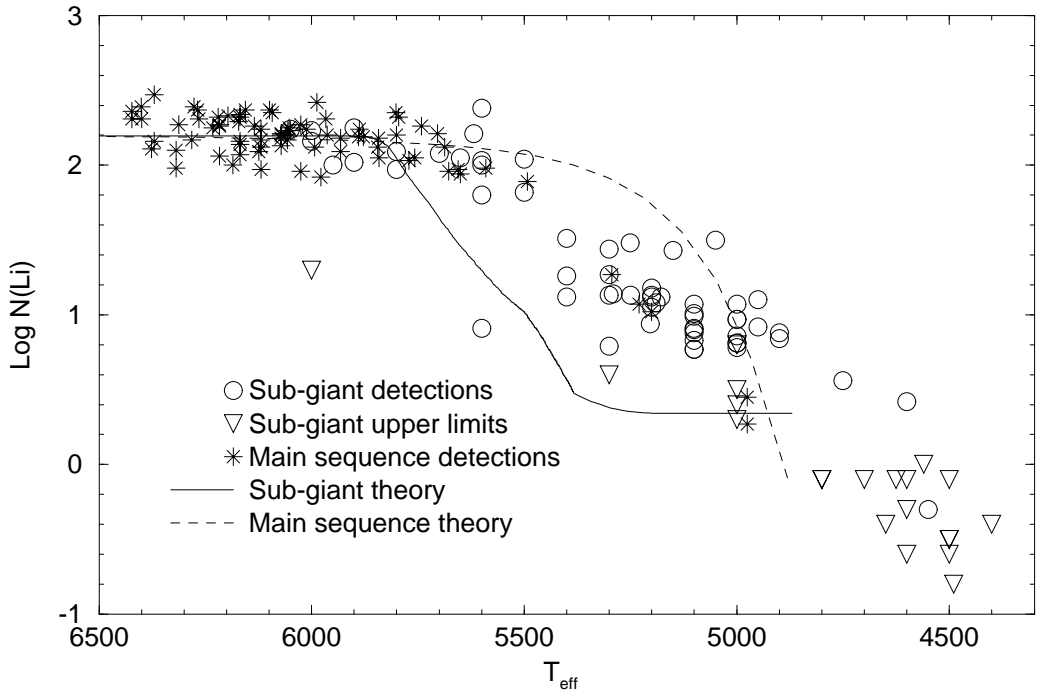

%\centerline{\psfig{figure=halolit.eps,height=9.0cm}}
\vspace*{9cm}
\caption
{Comparison of theoretical models to main sequence (Thorburn 1994) and
sub-giant observations (Pilachowski \ea ~1993).}
\label{subgli}
\end{figure}

Once a star evolves off the main sequence, the convection zone deepens
and material which had previously destroyed its \li ~becomes part of
the convection zone.  This results in a dilution of the surface \li
{}~abundance for stars on the sub-giant and red giant branch.  Thus,
observations of \li ~in halo sub-giants can be used to test if
the convection zone depth in the models agrees with reality.
Pilachowski, Sneden \& Booth (1993) obtained an extensive data set
which is plotted in Fig.~\ref{subgli} along with the main sequence
observations of Thorburn (1994), and the theoretical predications.  As
was clear from Fig.~\ref{kurli}, the main sequence models predict too
little depletion at cool temperatures.  In contrast, the sub-giant
models {\em over}-deplete \li ~in the range $5000 - 5700\,$K.  This
suggests that the convection zone depth is incorrect in these
sub-giant models, though in an opposite sense to the problem which
occurs for the main sequence models.  This  points to the need
for improved physics in halo star stellar models.  In addition, we
note that the coolest sub-giants ($\teff < 4800\,$K) have depleted
considerably more \li ~than is predicted by the models.  This
suggests that some form of extra mixing occurs on the giant branch, a
point which is discussed in more detail in Sec.~3.

The near uniformity of the plateau star \li ~abundances, which is
predicted by the standard models, puts severe constraints on mixing in
stellar radiative zones.  For example, microscopic diffusion predicts
greater \li ~depletions at hotter effective temperatures.  This is
clearly ruled out by the observations (Chaboyer \& Demarque 1994), and
so either diffusion does not operate in halo stars, or some other
mechanism acts to counteract the effects of diffusion on the \li
{}~abundances. Given the increasingly strong evidence from
helioseismology that diffusion is occurring in the Sun
(eg.~Christensen-Dalsgaard, Proffitt \& Thompson 1993), this latter
approach has been investigated by two groups.  Vauclair \& Charbonnel
(1995) found that stellar winds of order $10^{-12.5}\,\msun {\rm
yr}^{-1}$ coupled with microscopic diffusion were able to naturally
explain the plateau star \li ~abundances, and implied a primordial \li
{}~abundance of $\nli = 2.5\pm 0.1$.  These models also predict a small
dispersion in \li ~abundance among the plateau stars, and suggest that
the severely depleted plateau stars had significantly higher mass loss
rates.  Chaboyer \& Demarque (1994) found that stellar models which
coupled mixing induced by rotational instabilities with microscopic
diffusion provided a reasonable match to the plateau star \li
{}~abundances, and predicted a primordial \li ~abundance of $\nli \simeq
3.1$.  These models also predict that a real dispersion in \li
{}~abundance should exist among the plateau stars.

Standard models predict that no dispersion exists in the \li
{}~abundances of the plateau stars. Thus, determining whether or not a
dispersion exists is a key test which differentiates standard models
from models which include mixing in stellar radiative regions and
hence, predict a primordial \li ~abundance which is higher than that
observed in halo stars.  It is clear from present observations that if
a dispersion in \li ~abundances exists, it must be rather small.
Analysis of the dispersion is complicated by the fact that the
conversion from measured equivalent widths to actual \li ~abundances
is a strong function of the effective temperature of the star.  In
addition, one has to accurately determine the error in each individual
observation.

Deliyannis, Pinsonneault \& Duncan (1993) presented an analysis of the
dispersion in the colour-equivalent width plane and found that a real
dispersion exists, at the $\sim 20\%$ level.  Much of the uncertainty
surrounding the determinations of the dispersion can be overcome by
observing stars (with the same instrumental setup) which have
identical colours, reddenings (hence effective temperature), and ages.
Deliyannis, Boesgaard \& King (1995) have recently performed such a
test, by observing 3 sub-giants in the globular cluster M92 with the
Keck telescope.  By observing stars in a globular cluster, they were
able to measure \li ~abundances in stars which have identical colours,
reddenings and ages.  Hence, any difference in the observed equivalent
width of \li ~is direct evidence of a dispersion in the \li
{}~abundances. They found that a dispersion in \li ~abundance does
indeed exist among the plateau stars, which is evidence that \li
{}~depletion has occurred among the plateau stars.  Note however,
that Deliyannis \ea ~(1995) observed sub-giant stars with $\teff \sim
5700\,$K, which is near the cool edge of the plateau.  It would be
best to obtain similar observations for hotter stars, to ensure that
these sub-giants have not suffered dilation of \li ~at the surface due
to the deepening of the surface convection zone.

\section{Mixing on the Red Giant Branch}
\begin{figure}[t]
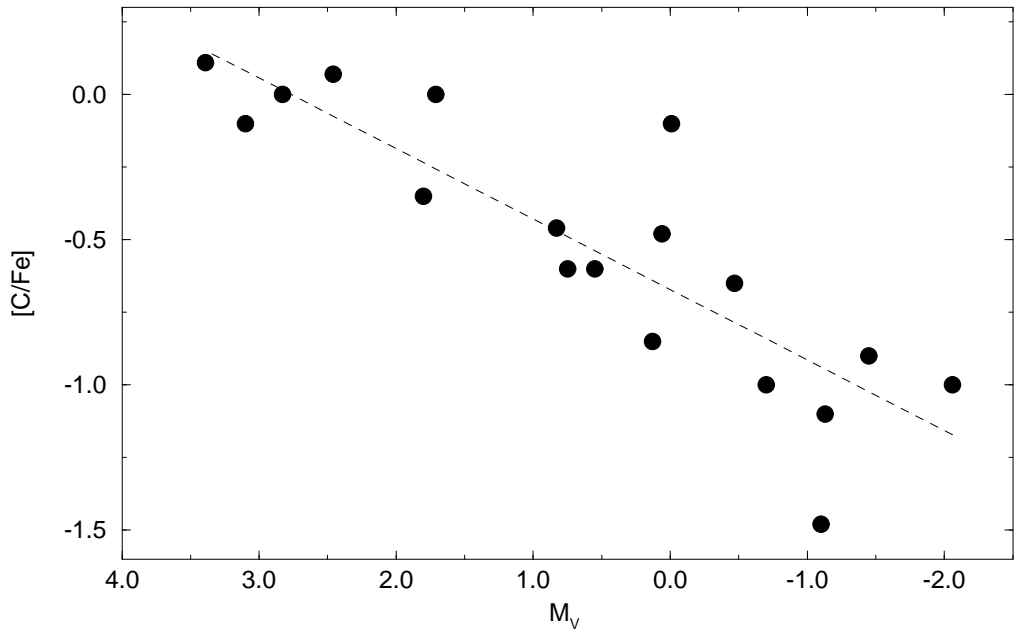

%\centerline{\psfig{figure=m92c.eps,height=9.0cm}}
\vspace*{9cm}
\caption
{Carbon abundances in M92 sub-giants and red giants as a function of
absolute magnitude, from Langer \ea ~(1986).  Only high quality
observations are plotted.  The dotted line is a simple least squares
fit to the data: ${\rm [C/Fe]} = -0.67 + 0.23\, {\rm M_V}$.}
\label{m92c}
\end{figure}

For a number of years, there has been evidence that CNO processed
material (which is enhanced in nitrogen, and depleted in carbon) is
being transported to the surface of red giant stars in the halo.
Perhaps the best evidence for this comes from the observations of
carbon in M92 giants by Langer \ea ~(1986).  They found that the
carbon to iron ratio ([C/Fe]) smoothly decreased from ${\rm [C/Fe]} =
+0.1$ around the main sequence turn-off to about ${\rm [C/Fe]} = -1.0$
for stars four magnitudes brighter than the turn-off, as shown in
Fig.~\ref{m92c}.   However,
standard stellar evolution models predict that the base of the
convection zone is never deep enough to dredge up CNO processed
material, and hence, [C/Fe] should remain constant.  The fact that the
decrease of [C/Fe] is smoothly related to the absolute magnitude of
the stars on the giant branch clearly points to the fact that some
non-standard, slow mixing is occurring in the radiative regions of
giant stars.  Further observational evidence has been found by Dickens
\ea ~(1991) who determined the abundances of carbon, nitrogen and
oxygen in giants in the globular cluster NGC 362.  They found that
[C/Fe] ranged from 0 to $-1$.  However, the total amount of C+N+O
(which is preserved in the CNO cycle) was constant.  Hence, the range
in the observed carbon abundances can only be explained by the fact
that CNO processed material has been brought to the surface in these
giant stars.

In addition to the above observations, I should note that there are
some observations of abundance patterns which are best explained by
primordial variations.  A recent review by Kraft (1994) provides a
detailed description of all of the abundance anomalies found by
observers in metal-poor giant stars. For example, there is a global
anti-correlation between oxygen and sodium abundances in globular
cluster red giants (Kraft \ea ~1993).  The most obvious explanation
for such a anti-correlation is due to primordial variations in oxygen
and sodium.  However, Denissenkov \& Denisenkova (1990) have suggested
that sodium could be produced via proton capture on neon in the CNO
burning region of a red giant star.  Thus, the global oxygen-sodium
anti-correlation could be further evidence that CNO processed material
is being transported to the surface of red giant stars.  This
possibility has been studied by Langer \ea ~(1993) and Denissenkov \&
Weiss (1995) who utilized an {\em ad hoc} slow mixing prescription
coupled with a detailed nuclear reaction network to predict changes in the
surface abundances as a star evolved up the red giant branch.  Both
groups concluded that it is possible to reproduce the global oxygen-sodium
anti-correlation by assuming that CNO processed material is being
brought to the surface of red giant stars.

Given the large number of observations which clearly indicate that CNO
processed material is dredged up to the surface of giant stars, in
contradiction with standard stellar models, it is clear that the
theoretical models need to be improved.  These models must incorporate
some form of slow mixing in the radiative regions, which will
progressively bring more CNO processed material to the surface of the
star as it evolves up the giant branch.  To do this, the driving
mechanism for the mixing must first be identified.  Sweigart \& Mengel
(1979) proposed that meridional circulation (induced by rotation) was
responsible for the mixing between the base of the surface convection
zone, and the CNO processed material in giant branch stars.  The
models of Sweigart \& Mengel (1979) were able to reproduce (and
indeed, predict) most of the observations provided that (i) the
convective envelope does not rotate as a solid body and (ii)
substantial reservoirs of angular momentum exist buried inside
main-sequence stars.  This first assumption appears to be reasonable,
given that the convective turnover times are long compared to the size
of the convection zone and the rotation velocities.  However, the
second assumption clearly needs to be explored in more detail.
Indeed, there has been little theoretical work on mixing in the
radiative regions of giant branch stars in recent years.  In view of
the wealth of observational evidence which is now available, this is
an area which clearly requires renewed attention from stellar
evolution theorists.

\section{Isochrones and Luminosity Functions}
\begin{figure}[t]
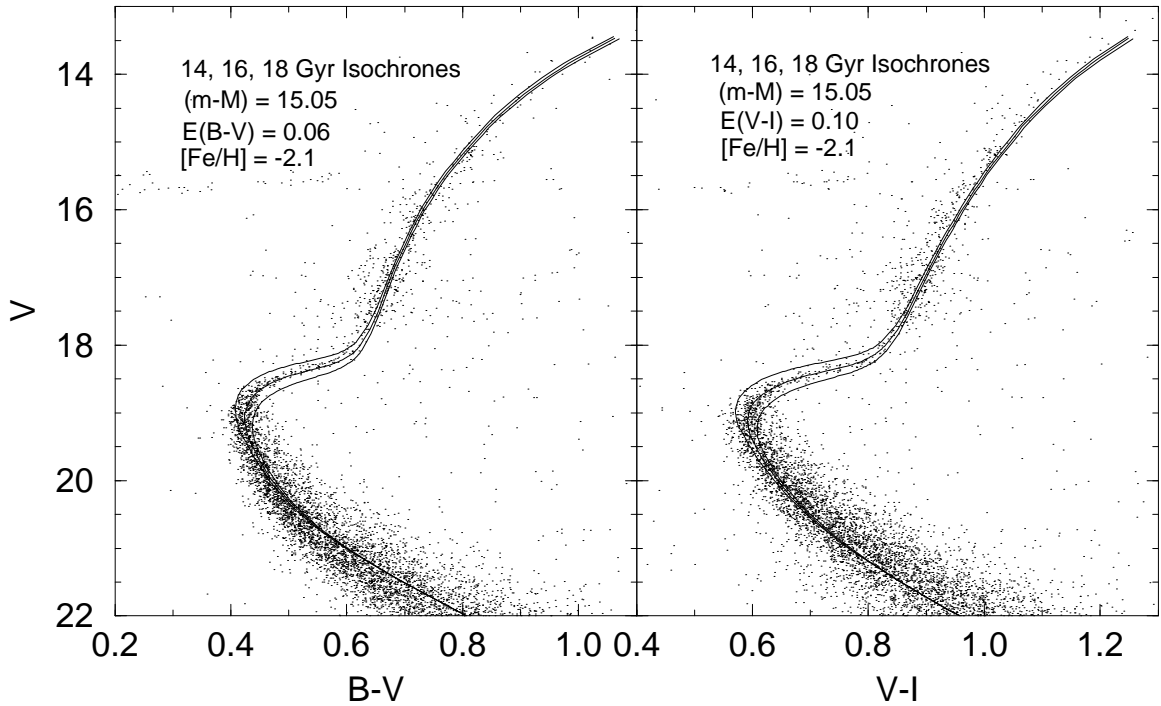

%\centerline{\psfig{figure=m68.eps,height=10.0cm}}
\vspace*{10cm}
\caption
{Comparison of theoretical isochrones to observations of M68 by Walker
(1994). A reasonable fit is obtained in both $\rm B-V$ (left panel)
and $V-I$ without any {\it ad hoc} colour shifts being applied to the
isochrones.}
\label{cmd}
\end{figure}

The calculation of isochrones and luminosity functions play a vital
role in many areas of astrophysics. As such, they probably represent
the most widely-used output of stellar evolution calculations.
Luminosity functions and isochrones are used in a variety of
applications: estimating the ages of star clusters and individual
stars; as basic building blocks for stellar population modeling and
galactic evolution codes; and even to determine the Hubble constant,
via the surface brightness fluctuation technique.  Modern isochrones
provide a good fit to colour-magnitude diagrams in globular clusters.
A typical example is shown in Fig.~\ref{cmd}, which compares the
recent set of Yale isochrones (Chaboyer \ea ~1995) to observations of
M68 (Walker 1994) in both $B-V$ and $V-I$.  A good fit is obtained to
the 16 Gyr isochrone in both colours, without any {\it ad hoc} colour
shifts.  These isochrones use a solar calibrated mixing length
($\alpha = 1.7$); if the mixing is changed by as little as 0.2 then it
is no longer possible to obtain a good fit.  While such good fits are
important to those who use isochrones for stellar population modeling,
the freedom to choose the mixing length imply that fitting theoretical
isochrones to observed colour-magnitude diagrams is not a good test of
the stellar evolution models.

\begin{figure}[t]
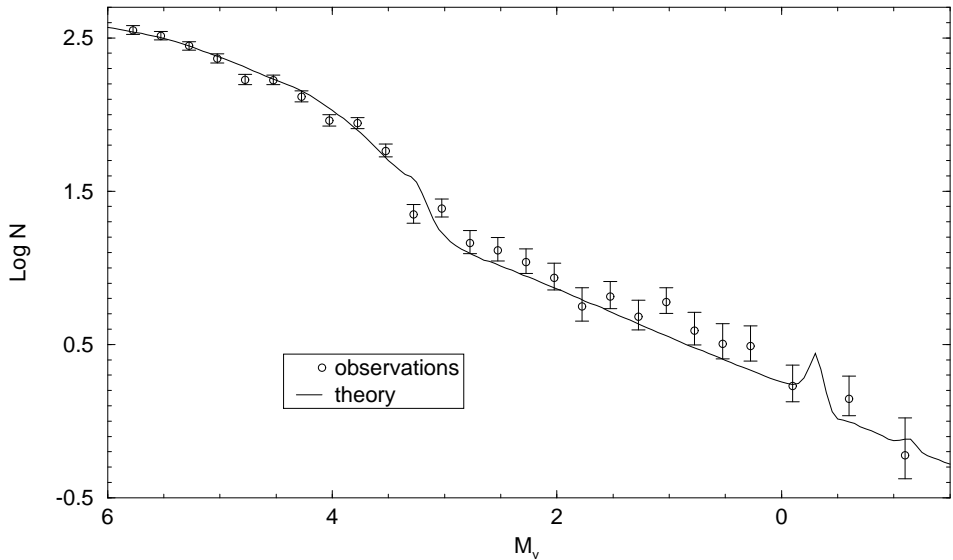

%\centerline{\psfig{figure=m30.eps,height=8.0cm}}
\vspace*{8cm}
\caption
{Comparison of a theoretical luminosity function from Chaboyer \ea
{}~(1995) to observations in M30 by Bolte (1994). The theoretical
luminosity function has been normalized to the main sequence ($5 \le
{\rm M_V} \le 6$) number counts.  Note the excess of observed stars
($\sim 15\%$) on the red giant branch (${\rm M_V} \la 1$) as compared
to the observations.}
\label{lf}
\end{figure}

A much more stringent test of stellar models is obtained by comparing
theoretical luminosity functions to observed number counts of stars in
globular clusters. If one focuses on the region from just below the
main sequence turn-off (within about 1 mag) to the tip of the giant
branch, the theoretical luminosity functions are independent of the
initial mass function and are only a function of the relative stellar
lifetimes in the various phases of evolution.  The stellar lifetimes
are a remarkably robust prediction of the theoretical models.  A
comparison between the Revised Yale isochrones (Green, Demarque \&
King 1987; based on stellar models calculated in the 1970's), and the
most recent Yale set reveals virtually no differences.  Unfortunately,
there are very few good quality observations of luminosity functions.
This is due to the fact that present CCDs have small physical sizes,
and the construction of luminosity functions requires good photometry
from below the main sequence turn-off to the tip of the giant branch,
over a large area of a globular cluster.  Although there were some
hints of discrepancy between the theoretical models and observations
in the past, in general the agreement was satisfactory.  However,
Bolte (1994) has recently published a very high quality luminosity
function of M30 ($\feh = -2.1$) which is clearly in disagreement with
theoretical calculations, as shown in Fig.~\ref{lf}. If the
theoretical luminosity functions are normalized to the main sequence
number counts, then there is about a 15\% excess in the number of
observed stars on the giant branch.  It is important to note that this
discrepancy is present for all reasonable choices of the assumed
distance modulus, reddening or metallicity.  In addition, as the
luminosity function is a rather robust prediction of the theoretical
models, it does not matter whose set of luminosity functions are used
to compare to the observations -- the discrepancy between theory and
observations is always present.  If the observations are interpreted
at face value, this suggests that the relative main
sequence/giant branch lifetimes are in error by about 15\%.  I caution
however, then M30 is a cluster which has undergone considerably
dynamical evolution, so it is important to have these observations
verified in other globular clusters.  Such observational efforts are
now underway by Peter Stetson and the Yale group.  If their results
confirm Bolte's (1994) observations, then a considerable revision in
the relative main sequence/giant branch lifetimes in halo stars will
be necessary.  The study of luminosity functions is clearly an area
worthy of further theoretical and observational effort.

\section{Globular Cluster Ages}
The determination of accurate globular cluster ages are important
in furthering our understanding of the formation of the Milky Way, and
in providing a lower limit to the age of the universe.  This latter
application has received increasing attention lately, as the rather old ages
inferred for globular clusters ($\sim 16$ Gyr) appear to be in
conflict with the young age inferred from measurements of the Hubble
constant, and the density of the universe (eg.~Freedman \ea ~1994).  If
the stellar ages are correct, then a rather dramatic revision of the
preferred cosmological model is in order (Leonard \& Lake 1995).  As
such, it is important for the stellar evolution theorist to provide a
reasonable estimate of the possible error in the determination of the
absolute ages of globular clusters.

Globular cluster ages are determined by comparing theoretical
isochrones to observed globular cluster colour magnitude diagrams.
There are a variety of ways in which this can be accomplished.  In
assessing the usefulness of the different methods, it is important to
recall that the colours predicted by the isochrones are subject to
large uncertainties due to our poor understanding of how to treat
convection in the superadiabatic regions near the surface, and due to
our limited knowledge of stellar atmospheres.  The $\Delta ({\rm
B-V})$ technique uses the difference in colour between the main
sequence turn-off and the base of the giant branch as an age
diagnostic (Sarajedini \& Demarque 1990; VandenBerg, Bolte \& Stetson
1990).  This technique is a sensitive test which can determine if age
differences exist between clusters of similar metallicity, but cannot
be relied upon to determine absolute ages, age differences between
clusters of different metallicity, or the absolute size of the
age difference for clusters of the same metallicity.

The most accurate theoretical ages are determined via methods which
rely on using the luminosity of the isochrones (and not colour) as the
age indicator.  A popular approach is to use the difference in
magnitude between the horizontal branch and the main sequence turn-off
(\dv).  In addition to being a luminosity based age determination,
this method has the advantage of being independent of reddening.
Unfortunately, the precise location of the main sequence turn-off, and
indeed the magnitude of the horizontal branch in the instability strip
can be difficult to determine observationally.  For this reason,
observational errors alone lead to typical $1\,\sigma$ errors of $\sim
10\%$ in the derived ages.  For the purposes of providing a reliable
estimate of the minimum age of the universe, this difficulty can be
overcome by determining the mean age of a number of globular clusters,
thereby minimizing the observational errors.

 From the theoretical point of view, determination of the absolute
magnitude of the horizontal branch is fraught with difficulties.
Horizontal branch stars are convective in the energy producing
regions, which leads to large uncertainties in the models.
Fortunately, there are several observationally-based techniques which
can be used to determine the absolute magnitude of RR Lyr stars
(\mvrr), and hence, the horizontal branch.  These calibrations can be
combined with theoretical isochrones which determine the absolute
magnitude of the turn-off to produce a semi-empirical calibration of
\dv ~as a function of age and metallicity.  The various observational
calibrations of \mvrr ~agree to within $\sim 0.25$ mag (for a
discussion of \mvrr ~see Carney, Storm \& Jones 1992; Chaboyer,
Demarque \& Sarajedini 1996).  This directly translates into an
overall error in the age determination of about 25\%.
\begin{table}[t]
  \begin{center}
  \begin{tabular}{ll}
\multicolumn{2}{c}{TABLE 1: GLOBULAR CLUSTER }\\
\multicolumn{2}{c}{AGES ERROR BUDGET}\\[3pt]
\hline\hline
{}~\\[-9pt]
\multicolumn{1}{c}{Description} &
\multicolumn{1}{c}{Error ($\pm \%$)}\\[2pt]
\hline
{}~\\[-8pt]
\mvrr & $ 13$\\
$[\alpha/{\rm Fe}]$~& $~7$\\[2pt]
$^4$He abundance & $~3$\\
Treatment of convection & $~9$\\
Diffusion & $~4$\\
Colour Transformation & $~3$\\
Nuclear Reaction Rates &$~3$\\
\hline
\end{tabular}
\end{center}
\end{table}

This large uncertainty in the derived ages has been widely discussed
in the literature (eg.~Renzini 1991).  However, not much attention has
been focused on the total error budget for the age determinations,
including possible errors due to the theoretical models.  I have
recently undertaken such an investigation (Chaboyer 1995).  The mean
age of the oldest globular clusters was determined using a number of
different isochrones, each of which were constructed using different
assumptions for the input physics.  I found that the opacities, model
atmospheres, convective overshoot, and uncertainties in the globular
cluster metallicity scale have little or no impact on the derived
ages.  A potentially large source of error involved the equation of
state.  Whether or not one included the Debye-H\"uckel correction into
the equation of state changed the ages by 7\%. Since then, I have
incorporated the OPAL equation of state tables (Rogers 1994), and
found that it gives gives nearly identical results to the equation of
state with the Debye-H\"uckel correction (Chaboyer \& Kim 1995).
Given the good agreement between the sophisticated OPAL equation of
state results, and the Debye-H\"uckel equation of state, I no longer
believe that the equation of state leads to significant uncertainties
in the derived ages. Using these new equation of state results,
combined with the ages published in Chaboyer (1995), my best estimate
for the mean age of the oldest globular clusters is 14.5 Gyr.  The
error budget associated with this number is tabulated in Table 1.
\begin{figure}[t]
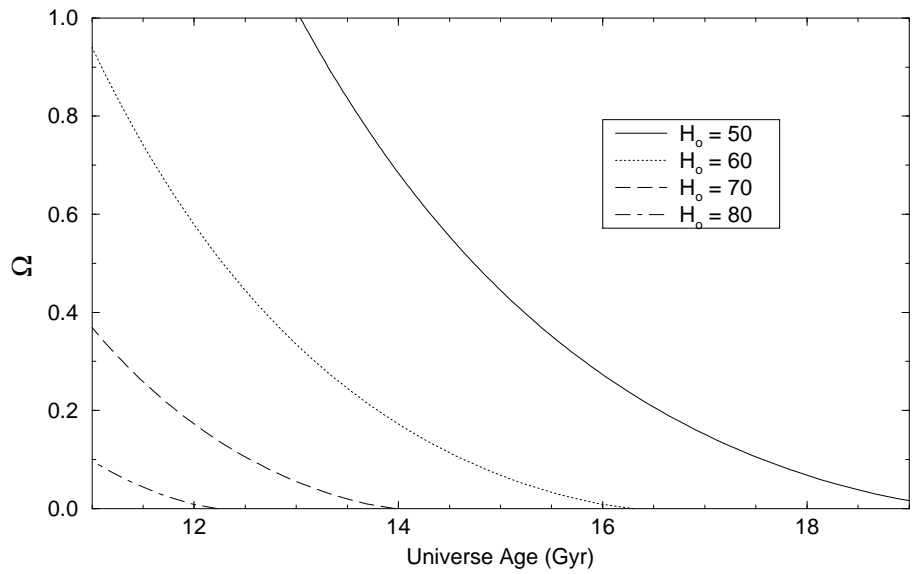

%\centerline{\psfig{figure=hoage.eps,height=8.0cm}}
\vspace*{8cm}
\caption
{The density of the universe $\Omega$ (in units of the critical
density) is plotted as a function of the age of the universe, for
various values of the Hubble constant (in units of km/s/Mpc), assuming
a zero cosmological constant. Current
estimates for the absolute ages of the globular clusters are in a
range 11 -- 18 Gyr, which provides a minimum estimate for the age of
the universe.}
\label{age}
\end{figure}

The percentage errors presented in Table 1, are {\em not} Gaussian
1-$\sigma$ error bars, but are meant to represent my best estimate of
the total possible error associated with a given uncertainty.  Next to
the uncertainty associated with \mvrr ~the largest single source of
error is the treatment of convection.  This is somewhat surprising, as
it is often assumed that the \dv ~technique yields ages which are
independent of the mixing length theory of convection used in stellar
models.  While it is true that the lifetime of the stars do not depend
on the mixing length, I found that changes in the mixing length
changed the shape of the isochrones around the turn-off, resulting in
different magnitudes for the turn-off (defined to be the bluest
point).  It may be possible to eliminate this uncertainty, by defining
the turn-off in terms of the luminosity function, as the number of
stars drop dramatically around the turn-off.  Another important source
of error in the age determination is the amount by which the
$\alpha$-capture elements (O, Mg, Si, S, and Ca) are enhanced over
their solar ratio.  Changing $[\alpha/{\rm Fe}]$ by 0.2 dex results in
a 7\% change in the derived age.  Given the possible errors listed in
Table 1, and recalling that my best estimate for the oldest globular
clusters are 14.5 Gyr old, it is reasonable to conclude that the true
age of the oldest globular clusters lies in the range 11 -- 18 Gyr.
The implications of this age estimate for cosmology are shown in
Fig.~\ref{age}, where the density of the universe $\Omega$, is plotted
as a function of age, for various values of the Hubble constant
(assuming no cosmological constant).  As there is general consensus
that observations require $\Omega \ga 0.3$, the minimum age of 11 Gyr
requires that a Hubble constant less than $\sim 70$ km/s/Mpc.  The
best estimate for the minimum age of the universe (14.5 Gyr) requires
that ${\rm H_o} \la 55$ km/s/Mpc.  Although there is still
considerable debate concerning a true value of the Hubble constant,
recent evidence appears to be favouring the higher values
(eg.~Freedman \ea ~1994; Riess, Press \& Kirshner 1995).  This suggests
that a non-zero cosmological constant is required (Leonard \& Lake
1995).

The rather large error in the globular cluster ages significantly
reduces the constraints imposed by the ages on cosmology.  Reducing
the size of this error will require an improved understanding of the
possible sources of error listed in Table 1.  Given the relative sizes
of the errors, effort should be concentrated on obtaining improved
estimates of \mvrr, $[\alpha/{\rm Fe}]$, and a better understanding of
how to treat convection in stars.  Better knowledge of diffusion,
nuclear reaction rates, and model atmospheres (used to derive the
colour transformation) are needed if one wishes to reduce the error in
the globular cluster age estimates below 10\%.

\section{Summary}
This review of the evolution of halo stars has highlighted some
inconsistencies between present stellar models and observations.
These inconsistencies merit further study.  As shown in
Fig.~\ref{subgli}, cool stellar models ($\teff \la 5600\,$K) which
utilize the best available input physics deplete too much \li ~on the
sub-giant branch, and too little \li ~on the main sequence.  This
implies that the convection zone depth is likely to be incorrect in
these models.  The plateau in \li ~abundances above $\teff \simeq
5700\,$K provides a stringent test for any possible mixing which
occurs in stellar radiative regions.  Standard stellar models are
unable to explain the $\sim 5\%$ of stars in the plateau region which
are severely depleted in \li.  In addition, there is evidence for a
dispersion in \li ~abundances for stars with identical effective
temperatures and ages (Deliyannis \ea ~1995). Such a dispersion is not
predicted by standard models, and is evidence that additional mixing,
and hence, \li ~depletion occurs in all halo stars.  Further evidence
that \li ~depletion has occurred in plateau stars is found via a
comparison to Pop I stars -- all of which show definite evidence for
depleting \li ~(Boesgaard 1991).  Stellar models which include
microscopic diffusion and a stellar wind (Vauclair \& Charbonnel 1995),
or diffusion and rotational mixing (Chaboyer \& Demarque 1994) do a
good job of reproducing the observed plateau star \li ~abundances and
suggest that the primordial \li ~abundance at least 0.3 dex higher
than what is presently observed in the plateau stars.

Observations of abundance anomalies in globular cluster red giant
stars have demonstrated for some time that some form of slow mixing
operates in the radiative regions of stars.  This evidence is plotted
in Fig.~\ref{m92c} which demonstrates that CNO processed material
(which is depleted in carbon) is slowly being brought to the surface
of red giant stars.  This is not predicted by standard stellar models.
Even though there has been a steady accumulation of observational
evidence demonstrating the failure of standard stellar models over the
last several years, there has been little theoretical work to improve
the stellar models.  Sweigart \& Mengel (1979) proposed that
meridional circulation (induced by rotation) was responsible for
mixing CNO processed material into the convection zone (and hence, to
the surface) in giant branch stars.  Their models were in reasonable
agreement with the observations, and should be pursued further in
light of the extensive observational database which is now available.

Theoretical isochrones which are calculated from modern stellar
evolution models provide a good match to observed colour-magnitude
diagrams (cf.~Fig.~\ref{cmd}).  However, there is some evidence that
the luminosity functions do not match the observations
(Fig.~\ref{lf}), predicting relative main sequence/giant branch
lifetimes which are in error by $\sim 15\%$.  This merits further
observational and theoretical study, for this may have important
implications for stellar evolution models.  Determining the age of the
oldest globular clusters provides the best estimate for the minimum
age of the universe, and is of great interest to cosmologists.
Currently, the best estimate yields an age of 14.5 Gyr for the oldest
globular clusters.  A critical evaluation of the possible error
associated with this estimate (Table 1, see Chaboyer 1995) suggests
that the true age of the oldest globular clusters lies in the range 11
-- 18 Gyr.  This rather large allowed range in the globular cluster
ages significantly reduces the constraints imposed on cosmology.
Reducing the error in the ages to below the 10\% level will require an
improved understanding of all of the possible error sources listed in
Table 1.

The study of the evolution of halo stars has an impact on many areas
of astrophysics, helping us to understand questions related to big
bang nucleosynthesis, cosmology, and galaxy formation.  The examples
highlighted in this review illustrate the fact that our knowledge of
the evolution of halo stars is incomplete.  Given the broad ranging
applications of halo star models, an improved knowledge of the evolution
of halo stars is highly desirable and worthy of further study.

%
% USE A SECTION WITHOUT NUMBER FOR THE ACKNOWLEDGEMENTS
%
\section*{Acknowledgements}
I would like to thank H.~Craig for her careful editing of this manuscript.
%
% BEGIN THE REFERENCE LIST WITH \beginrefer
% USE \refer BEFORE THE REFERENCES AND BEGIN A NEW PARAGRAPH AFTER THE
% REFERENCE !
% TYPESET THE NAME OF THE JOURNAL IN \sl
% TYPESET THE NUMBER OF THE JOURNALE IN \bf
% DO NOT FORGET TO END THE LIST WITH \endrefer
%

\beginrefer

\refer Boesgaard, A.M.~1991, {\sl ApJ}, {\bf 370}, L95

\refer Bolte, M.~1994, {\sl ApJ}, {\bf 431}, 115

\refer Carney, B.W.,~Storm, J.~\& Jones, R.V.~1992, {\sl ApJ}, {\bf 386}, 663

\refer Castellani, V.~\& Santis, R.D.~1994, {\sl ApJ}, {\bf 430}, 624

\refer Chaboyer, B.~1995, {\sl ApJ}, {\bf 444}, L9

\refer Chaboyer, B.~\& Demarque, P.~1994, {\sl ApJ}, {\bf 433}, 510

\refer Chaboyer, B., Demarque, P., Guenther, D.B., Pinsonneault,
M.H.~\& Pinsonneault, L.L. 1995, in `The Formation of the
Milky Way', eds. E.J. Alfaro \& G. Tenorio-Tagle (Cambridge: Cambridge
University Press), in press

\refer Chaboyer, B., Demarque, P.~\& Sarajedini, A.~1996, {\sl ApJ},
in press

\refer Chaboyer, B.~\& Kim, Y.-C.~1995, {\sl ApJ}, in press (Dec.~1/95)

\refer Christensen-Dalsgaard, J., Proffitt, C.R.~\& Thompson,
M.J.~1993, ApJ, 403, L75

\refer Deliyannis, C.P., Boesgaard, A.M.~\& King, J.R. 1995, {\sl
ApJL}, in press

\refer Deliyannis, C.P., Pinsonneault, M.H.~\& Duncan, D.K.~1991, {\sl
ApJ}, {\bf 414}, 740

\refer Denissenkov, P.A.  \& Denisenkova, S.N.~1990, {\sl Soviet
Astron.~Lett.\/}, {\bf 16}, 275

\refer Denissenkov, P.A.  \& Weiss, A.~1995, {\sl A\&A}, submitted

\refer Dickens, R.J., Croke, B.F.W, Cannon, R.D.~\& Bell, R.A.~1991,
{\sl Nature}, {\bf 351}, 212

\refer Freedman, W. L., Madore, B. F., Mould, J. R., Hill, R., Ferrarese, L.,
Kennicutt, R. C., Saha, A., Stetson, P. B., Graham, J. A., Ford, H.,
Hoessel, J. G., Huchra, J., Hughes, S. M.~\& Illingworth, G. D.~1994,
{\sl Nature}, {\bf 371}, 757

\refer Green, E.M., Demarque, P.~\& King, C.R.~1987, `The Revised Yale
Isochrones and Luminosity Functions' (Yale Univ. Obs., New Haven)

\refer Kraft, R.P.~1994, {\sl PASP}, 106, 553

\refer Kraft, R.P., Sneden, C., Lander, G.E.~\& Shetrone, M.D.~1993,
{\sl AJ}, {\bf 106}, 1490

\refer Kurucz, R.L.~1992, in IAU Symp. 149, The Stellar
Populations of Galaxies, ed. B. Barbuy, A. Renzini, (Dordrecht: Kluwer), 225

\refer Langer, G.E., Kraft, R.P., Carbon, D.F., Friel, E.~\& Oke,
J.B.~1986, {\sl PASP}, {\bf 98}, 473

\refer Langer, G.E., Bolte, M., Prosser, C.F.~\& Sneden, C.~1993, {\sl
ASP Conf.~Ser.}, {\bf 48}, 206

\refer Leonard, S.~\& Lake, K.~1995, {\sl ApJ}, {\bf 441}, L55

\refer Pilachowski, C.A., Sneden, C.~\& Booth, J. {\sl ApJ}, {\bf
407}, 699

\refer Renzini, A.~1991, in `Observational Tests of Cosmological
Inflation', eds.~T.~Shanks, \ea (Dordrecht: Kluwer), 131

\refer Riess, A.G., Press, W.H.~\& Kirshner, R.P.~1995, {\sl ApJ},
{\bf 445}, L91

\refer Rogers, F.J.~1994, in `The Equation of State in Astropyhsics', IAU
Coll.~147, ed.~G. Chabrier and E.~Schatzman (Cambridge:
Cambridge University Press), 16

\refer Sarajedini, A.~\& Demarque, P.~1990, {\sl ApJ}, {\bf 365}, 219

\refer Smith, M.S., Kawano, L.H. \& Malaney, R.A.~1993, {\sl ApJS},
{\bf 85}, 219

\refer Smith, V.V. Lambert, D.L. \& Nissen, P.E.~1993, {\sl ApJ},
{\bf 408}, 262

\refer Sweigart, A.V.~\& Mengel, J.G.~1979, {\sl ApJ}, {\bf 229}, 624

\refer Spite, F. \& Spite, M.~1982, {\sl A\&A}, {\bf 115}, 357

\refer Thorburn, J.A.~1994, {\sl ApJ}, {\bf 421}, 318

\refer Vauclair, S.~\& Charbonnel, C.~1995, {\sl A\&A}, {\bf 295}, 715

\refer Walker, A.R.~1994, {\sl AJ}, {\bf 108}, 555

\refer Walker, T.P., Steigman, G., Kang, H., Schramm, D.M.~\& Olive,
K.A.~1991, {\sl ApJ}, {\bf 376}, 51

\refer VandenBerg, D.A., Bolte, M.~\& Stetson, P.B.~1990, {\sl AJ},
{\bf 100}, 445

\endrefer

\section*{Discussion}
\noindent
{\bf I.~Roxburgh:}~~
If the inner 30\% or so of the mass of low mass
stars was mixed on the main sequence, would one predict results that
were in conflict with observations (or more in conflict than standard
models)?

\vspace{4pt}
\noindent
{\bf B.~Chaboyer:}~~
Mixing the inner part of a halo star is essentially assuming that such
stars have convective cores.  Stars which have convective cores
display a characteristic hook in colour-magnitude diagram as they
leave the main sequence.  These hooks are observed in colour-magnitude
diagrams of young star clusters, as predicted by stellar
models. However, globular cluster colour-magnitude diagrams contain no
such hooks, which is a strong argument against substantial mixing
occurring in the inner region of low mass stars.

\vspace{7pt}
\noindent
{\bf B.~Dorman:}~~
I would like to comment that the zero-point of the \mvrr ~relation is
not as clear-cut as you seem to paint it.  The evolutionary models
with $Y=0.23$ (which give the bright zero-point) are consistent with
primordial helium, and R-method determination of the helium abundance
in globular clusters.  The brighter Walker (1992) zero-point is
apparently supported also by recent HST observations of M31 clusters
(Faber 1995, private communication).  It is supported by Carney \ea
{}~(1992) re-analysis of the main-sequence fitting technique.  Further,
the Baade-Wesselink analysis may be subject to a systematic
underestimate in the radius (eg.~Castellani \& Santis 1994).
Statistical parallax methods that are independent of the
Baade-Wesselink distances give results (Barnes \& Hawley 1986) that
are consistent within the errors with the brighter distance scale.

\vspace{4pt}
\noindent
{\bf B.~Chaboyer:}~~
I agree with you that the question of the zero-point in the \mvrr
{}~relation is still a matter of debate.  As such, I have assumed an
uncertainty in the zero-point of 0.25 mag, which is the typical
difference between the `bright' and `faint' calibrations of the \mvrr
{}~zero-point.  I am not aware of the HST M31 observations, but I should
point out that Walker's (1992) value of the zero-point was based on an
LMC distance modulus of 18.5 mag.  Gould (1995) has re-analyzed the
distance to the LMC based on the rings surrounding SN1987A, and
determined a lower limit to the distance modulus which is 18.37 mag.
This makes Walker's zero-point {\em fainter} by 0.13 mag, removing
much of the discrepancy with the statistical parallax results.  My
understanding of the Carney \ea ~(1992) main-sequence fitting results
is rather different than yours. I believe that they determined a
zero-point which was even fainter than the statistical parallax work,
and incompatible with Walker's (1992) value.  I agree that the
Baade-Wesselink analysis may be subject to systematic uncertainties
and is best used to determine the slope of the \mvrr ~relation with
metallicity, and not the zero-point.

\vspace{7pt}
\noindent
{\bf C.~Charbonnel:}~~
In our models (Vauclair \& Charbonnel 1995) which include microscopic
diffusion and stellar wind, we not only reproduce the slope of the
lithium `plateau' for halo stars, but, with a reasonable range of mass
loss rates, we also reproduce the observed dispersion and explain the
upper limits seen well below the plateau.  So my question is: What is
the primordial abundance that you predict with rotation-induced
mixing, and how do you explain all of the characteristics of the
plateau (slope, dispersion and upper limits) without {\em ad hoc}
assumptions?

\vspace{4pt}
\noindent
{\bf B.~Chaboyer:}~~
Our models with rotation-induced mixing and microscopic diffusion
predict a primordial abundance of $^7$Li which is 1 dex above the
presently observed abundance.  However, given the uncertainties in the
models, I believe that a more accurate statement is that the $^7$Li in
plateau stars has been depleted by at least 0.4 dex, implying a
primordial abundance of $\nli > 2.6$.  These models with
rotation-induced mixing and microscopic diffusion are able to
naturally explain the slope of the plateau, using the same parameters
which are able to match $^7$Li observations in the open clusters and
the Sun.  Our models account for the dispersion in the plateau if we
assume the majority of stars are born as slow rotators ($\sim
10\,$km/s), with about $\sim 20\%$ having rotation
velocities $30\,$km/s or greater.  This implied distribution in
initial rotation velocities is similar to what is observed in present
day T-Tauri stars.  The very low upper limits observed in about 5\% of
the halo star sample are more difficult to reconcile with our models.
Perhaps they are a sign that a few stars were born with very high
rotation velocities, or perhaps they are due to high mass loss rates as you
suggest.

\end{document}